\documentclass{aa}
\usepackage{graphicx}
\usepackage{times}
\begin{document}

\def\gtsim{\raisebox{-.5ex}{$\;\stackrel{>}{\sim}\;$}}
\newcommand{\mrm}[1]{\mathrm{#1}}
\newcommand{\dmrm}[1]{_{\mathrm{#1}}}
\newcommand{\umrm}[1]{^{\mathrm{#1}}}

\thesaurus{03(11.01.2; 11.10.1; 11.17.4 3C 273; 13.18.1)}

\title{Modelling the submillimetre-to-radio flaring behaviour of 3C~273}
\author{
M. T\"urler \inst{1,2} \and
T.J.-L. Courvoisier \inst{1,2} \and
S. Paltani \inst{1,2}
}
\institute{
Geneva Observatory, ch. des Maillettes 51, CH-1290 Sauverny, Switzerland \and
\textit{INTEGRAL} Science Data Centre, ch. d'\'Ecogia 16, CH-1290 Versoix, Switzerland
}
\offprints{M. T\"urler (ISDC)}
\mail{Marc.Turler@obs.unige.ch}
\date{Received date / Accepted date}
\maketitle

\begin{abstract}
We present a new approach to derive the observed properties of synchrotron outbursts in relativistic jets.
The idea is to use the very well sampled submillimetre-to-radio long-term light curves of 3C~273 to extract the spectral and temporal evolution of a typical outburst.
The method consists in a decomposition of these light curves into a series of twelve self-similar flares.
With a model-independent parameterization, we find that the obtained outburst's evolution is in good qualitative agreement with the expectations of shock models in relativistic jets.
We then derive, by a second approach, the relevant parameters of three-stage shock models.

We observe for the first time that the optically thin spectral index is steeper during the initial rising phase of the evolution than during the final declining phase as expected by the shock model of Marscher \& Gear (\cite{MG85}).
We obtain that this index flattens from $\alpha\!=\!-1.1$ ($S_{\nu}\!\propto\!\nu^{\alpha}$) to $\alpha\!=\!-0.5$, in good agreement with what is expected from a power law electron energy distribution of the form $N(E)\!\propto\!E^{-2.0}$.
The observed flattening gives support to the idea that radiative (synchrotron and/or Compton) losses are the dominant cooling process of the electrons during the initial phase of the shock evolution.

Two other results give us confidence in our decomposition:
1) the outbursts that we identify do well correspond to the VLBI components observed in the jet and 2) there is strong evidence that high-frequency peaking outbursts evolve faster than low-frequency peaking outbursts.
We propose that this last correlation is related to the distance from the core of the jet at which the shock forms.

\end{abstract}

\section{Introduction}
\label{introduction}

Submillimetre-to-radio light curves of blazars show evidence of prominent structures, or flares, apparently propagating from high to low frequencies.
A decisive step in the understanding of these flares was done by Marscher \& Gear (\cite{MG85}, hereafter MG85).
They studied the strong 1983 outburst of \object{3C~273} by constructing at two epochs a quasi-simultaneous millimetre-to-infrared spectrum after subtracting a quiescent emission assumed to vary on a much longer time scale.
They successfully fitted these two flaring spectra with self-absorbed synchrotron emission and showed that their temporal evolution can be understood as being due to a shock wave propagating down a relativistic jet.
They identified three stages of the evolution of the shock according to the dominant cooling process of the electrons: 1) the Compton scattering loss phase, 2) the synchrotron radiation loss phase and 3) the adiabatic expansion loss phase.

Another shock model was developed by Hughes et al. (\cite{HAA85}) simultaneously to that of MG85.
Their piston-driven shock model reproduces well the lower frequency flux and polarization observations of outbursts in \object{BL Lacertae}, but fails to describe the observed behaviour in the millimetre domain.
A generalization of the three-stage shock model of MG85 was presented by Valtaoja et al. (\cite{VTU92}).
Their model, based on observations, describes qualitatively the three stages of the MG85 model without going into the details of the physics of the shock.
Finally, Qian et al. (\cite{QWB96}) proposed a burst-injection model to study the spectral evolution of superluminal radio knots.
Their theoretical calculation is able to reproduce well the observed spectral evolution of the C4 knot in \object{3C~345} (Qian \cite{Q96}).

To constrain these shock models, we need to extract the properties of the outbursts from the observations.
This step is difficult both at high and at low frequencies.
At high frequencies because of the brevity of the outbursts that last only a few days to months, thus requiring a very well sampled set of observations in the not easily accessible submillimetre spectral range.
At radio frequencies because they very often overlap due to their longer duration, making it difficult to isolate them.

The best observational constraints for the model of MG85 were obtained by Litchfield et al. (\cite{LSR95}) for the blazar \object{3C~279} and by Stevens et al. (\cite{SLR95}, \cite{SLR96}, \cite{SRG98}) for \object{PKS~0420$-$014}, 3C~345 and 3C~273, respectively.
All these studies are based on isolated outbursts.
The method used consists in constructing simultaneous multi-frequency spectra for as many epochs as possible after the subtraction of a quiescent spectrum assumed to be constant with time.
The subtraction of a quiescent spectrum is convenient and seems to give good results, but has only weak physical justification.
In 3C~273, there was a period of nearly constant flux at millimetre frequencies lasting just more than one year in 1989--1990, which was interpreted as its quiescent state (Robson et al. \cite{RLG93}).
At radio frequencies, however, no similar constant flux period was ever observed and there is no evidence that such a state exists at a significant level above the contribution of the jet's hot spot \object{3C~273A} (see Fig.~2 of T\"urler et al. \cite{TPC99}).

The different approach presented here to derive the observed properties of the outbursts has the advantage to not rely on the assumption of a quiescent emission.
The idea is to decompose a set of light curves covering a large time span into a series of flares.
To our knowledge, the first attempt of such a decomposition was made by Legg (\cite{L84}), who fitted a ten years radio light curve of \object{3C~120} with twelve self-similar outbursts.
Recently, Valtaoja et al. (\cite{VLT99}) decomposed the 22\,GHz and 37\,GHz radio light curves of many active galactic nuclei into several exponentially rising and decaying outbursts.
What is new in our approach is that we fit the same outbursts simultaneously to twelve light curves covering more than two decades of frequency from the submillimetre to the radio domain.
This adds a new dimension to the decomposition: the evolution of a flare is now a function of both time and frequency.
The aim is to obtain both the spectral and temporal properties of a typical flare, from which individual flares differ only by a few parameters.

We use the light curves of 3C~273, the best observed quasar, to have as many observational constraints as possible.
The flaring behaviour of 3C~273 was already the subject of several previous studies (e.g. Robson et al. \cite{RLG93}; Stevens et al. \cite{SRG98}).
Stevens et al. (\cite{SRG98}) obtain results for the first stage of the strong 1995 flare in very good agreement with the predictions of the MG85 shock model.
The new approach presented here is however more powerful to constrain the two following stages of the evolution.

We describe below two different approaches.
In Sect.~\ref{approach1} we model the light curve of each outburst by an analytic function that can smoothly evolve with frequency, whereas in Sect.~\ref{approach2} we directly model a self-absorbed synchrotron spectrum that evolves with time.
The first approach is easier to implement, since it allows us to begin the decomposition with a single light curve before adding the others progressively.
The second approach is more physical and gives better constraints to shock models.
Our results are discussed in Sect.~\ref{discussion} and summarized in Sect.~\ref{summary}.

Throughout this paper the frequency $\nu$ is as measured in the observer's frame and ``$\log$'' refers to the decimal logarithm ``$\log_{10}$''.
The convention for the spectral index $\alpha$ is $S_{\nu}\!\propto\!\nu^{+\alpha}$.

\section{Observational material}
\label{material}

This study is based on the light curves of the multi-wavelength database of 3C~273 presented by T\"urler et al. (\cite{TPC99}).
The twelve light curves we use are the five radio light curves: 5\,GHz, 8.0\,GHz, 15\,GHz, 22\,GHz and 37\,GHz and the seven millimetre/submillimetre (mm/submm) light curves: 3.3\,mm, 2.0\,mm, 1.3\,mm, 1.1\,mm, 0.8\,mm, 0.45\,mm and 0.35\,mm.
At low frequency (5 to 15\,GHz), we consider only the measurements of the University of Michigan Radio Astronomy Observatory (UMRAO).
The observations at 22\,GHz and 37\,GHz are mainly from the Mets\"ahovi Radio Observatory in Finland.
The mm/submm observations are from various sources including the James Clerk Maxwell Telescope (JCMT), the Swedish-ESO Submillimetre Telescope (SEST) and the ``Institut de Radio-Astronomie Millim\'etrique'' (IRAM).

We analyse the observations from 1979.0 to 1996.6, except at low frequency where we extend the analysis up to: 1997.2 (15\,GHz), 1997.5 (8.0\,GHz) and 1998.0 (5\,GHz), in order to include the decay of the 1995 flare.
In the mm/submm range, we average repeated observations made within 3 days to avoid oversampling of the light curves at some epochs.
This leaves us a total of 4352 observational points to constrain our fits.
To observations without known uncertainties, we assign the average uncertainty of the other observations at the same frequency.
The light curves are treated as if all their observations were made exactly at the same frequency, i.e. small differences of the observing frequency from one measurement to the other are not taken into account.
This simplification should not much affect the results, since the spectrum is rather flat ($\alpha\gtsim -0.5$) in the considered submillimetre-to-radio domain (e.g. T\"urler et al. \cite{TPC99}).

\section{The light-curve approach}
\label{approach1}

We describe here an approach in which we minimize the number of model-dependent constraints.
The light curve of each outburst at a given frequency is described by a simple analytical function.
The choice of this function is purely empirical and does not rely on any physical model.
The evolution with frequency of the outburst's light curve is left as free as possible.
This model has therefore many free parameters, which can adapt to a wide range of different situations.

\subsection{Number of outbursts}
\label{number}

One crucial parameter of the decomposition is the number of outbursts.
Pushed by the wish to reproduce the small features seen in the light curves, one is tempted to add always more outbursts to the fit.
In T\"urler et al. (\cite{TCP99}), we published the results of a decomposition into nineteen outbursts using an approach which is similar to that described below.
Here, we try to minimize as much as possible the number of outbursts to better constrain their spectral and temporal evolution.
We end with twelve flares, which are absolutely necessary to describe the main features of the light curves.

The aim of the decomposition is not to reproduce the detailed structure of the light curves, but to derive the main characteristics of the outbursts.
As a consequence, the $\chi^2$ that we shall obtain will be statistically completely unacceptable and will have no meaning in terms of the probability that the model corresponds to what is observed.
We will however refer to the obtained values of the reduced $\chi^2$ (cf. Sect.~\ref{results1}), because it is the usual way to express the quality of a fit.

\begin{figure}[tb]
\includegraphics[width=\hsize]{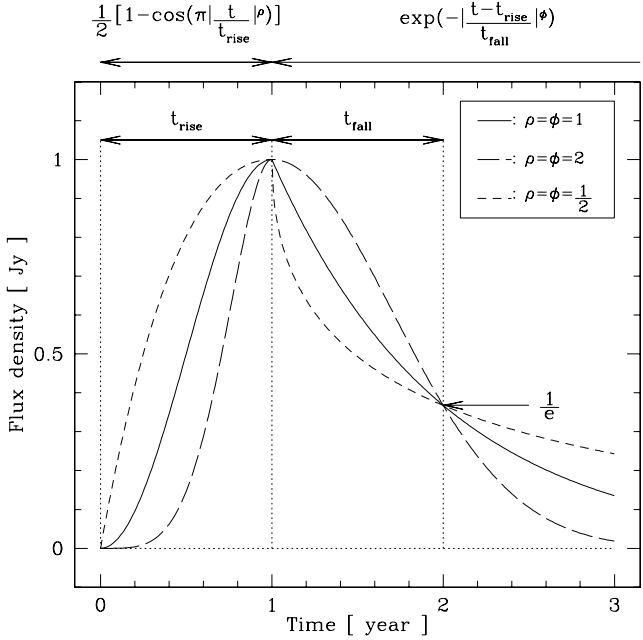}
\caption{
Model light curve of an outburst defined by Eqs.~(\ref{rise}) and (\ref{fall}), which starts at time $t_0\!=\!0$ and peaks at an amplitude of $A(\nu)\!=\!1$\,Jy.
The three different line types show the effect of varying $\rho(\nu)$ and $\phi(\nu)$
\label{func}
}
\end{figure}

\subsection{Parameterization}
\label{model1}

At a given frequency $\nu$, we model the light curve $S_{\nu}(t)$ of a single outburst of amplitude $A(\nu)$, starting at time $t\!=\!t_0$ and peaking at $t=t_0+t\dmrm{rise}(\nu)$ by
\begin{equation}
\label{rise}
S_{\nu}(t)=\frac{A(\nu)}{2}\left[ 1-\cos{\left( \pi\left( \frac{t-t_0}{t\dmrm{rise}(\nu)} \right) ^{\rho(\nu)}\right)}\right],
\end{equation}
if $t_0\leq t < t_0 + t\dmrm{rise}(\nu)$ and by
\begin{equation}
\label{fall}
S_{\nu}(t)=A(\nu)\exp{\left(-\left(\frac{t\!-\!t_0\!-\!t\dmrm{rise}(\nu)}{t\dmrm{fall}(\nu)} \right) ^{\phi(\nu)}\right) },
\end{equation}
if $t\geq t_0 + t\dmrm{rise}(\nu)$.
The exponents $\rho(\nu)$ and $\phi(\nu)$ define the shape of the light curve at frequency $\nu$ and $t\dmrm{fall}(\nu)$ is the $\mrm{e}$-folding decay time of the flare at frequency $\nu$.
Different time profiles of an outburst defined by Eqs.~(\ref{rise}) and (\ref{fall}) are shown in Fig.~\ref{func}.

Rather than constraining the outburst parameters ($A(\nu)$, $t\dmrm{rise}(\nu)$, $t\dmrm{fall}(\nu)$, $\rho(\nu)$ and $\phi(\nu)$) at each of the twelve light curve's frequencies, we describe their logarithm by a cubic spline which we parameterize at only four frequencies spaced by 0.75\,dex and covering the 3\,--\,600\,GHz range (see Fig.~\ref{param}).
This reduces the number of free parameters by a factor three, while keeping the parameterization completely model-independent.
We thus need a total of $5\!\times\!4$ parameters to fully characterize the spectral and temporal evolution of an outburst, i.e. a surface in the three dimensional $(S,\nu,t)$-space (cf. Fig.~\ref{evo1}).

We impose that all individual outbursts are self-similar, in the sense that they all have the same evolution pattern, i.e. the same shape of the surface in the $(S,\nu,t)$-space.
What we allow to change from one outburst to the other is the normalization in flux $S$, frequency $\nu$ and time $t$, which changes, respectively, the amplitude of the outburst (strong or weak), the frequency at which the emission peaks (high- or low-frequency peaking) and the time scale of the evolution (long-lived or short-lived).
A change in normalization corresponds to a shift of the position of the outburst's characteristic surface in the $(\log{S},\log{\nu},\log{t})$-space.
To define this position, we take the point of maximum flux as an arbitrary reference point on the surface.
On average among all individual outbursts, this point is located at $(\langle\log{S}\rangle,\langle\log{\nu}\rangle,\langle\log{t}\rangle)$ and this average normalization defines what we call the \textit{typical outburst} of 3C~273.
We denote by $\Delta\log{S}$, $\Delta\log{\nu}$ and $\Delta\log{t}$ the logarithmic shifts of this point with respect to the average position, i.e.
$\Delta\log{k}\!=\!\log{k}-\langle\log{k}\rangle,\,\forall\,k\!=\!S,\,\nu,\,t$.
These $12\!\times\!3$ logarithmic shifts plus the $12$ different start times $t_0$ of the flares give a total of $48$ parameters used to define the specificity of all outbursts.

The superimposed decays of the outbursts that started before 1979 are simply modelled by an hypothetical event of amplitude $A_0(\nu)$ at time $t=t_0+t\dmrm{rise}(\nu)=1979.0$ and decaying with the $\mrm{e}$-folding time $t\dmrm{fall}(\nu)$ of the typical outburst at frequency $\nu$.
The variation of the amplitude $A_0(\nu)$ with frequency is modelled by a cubic spline as for the five other variables, but parameterized at four slightly lower frequencies ($\log{(\nu/\mbox{GHz})}=$ 0.5, 1.0, 1.5 and 2.0), due to the fact that $A_0(\nu)$ is only well constrained for the radio light curves.
Finally, we assume a constant contribution to the light curves due to the quiescent emission of the jet's hot spot 3C~273A.
This emission is modelled with a power law spectrum as given in T\"urler et al. (\cite{TPC99}).

To summarize, this first parameterization uses a total of $72$ ($20+48+4$) parameters to adjust the 4352 observational points in the twelve light curves.
The great number of free parameters still leaves more than four thousand degrees of freedom (d.o.f.) to the fit.
The simultaneous fitting of the twelve light curves is performed by many iterative fits of small subsets of the $72$ parameters.

\begin{figure}[tb]
\includegraphics[width=\hsize]{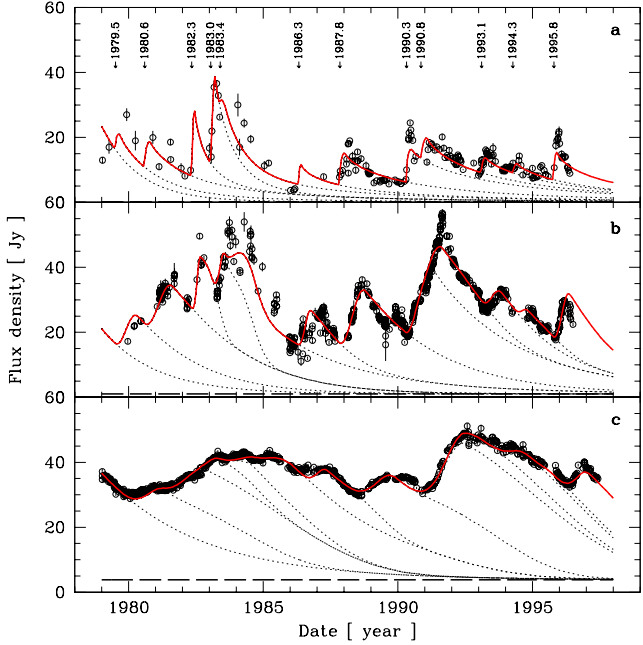}
\caption{
\textbf{a--c.} Three of twelve light curves fitted by the cumulative sum (solid line) of twelve outbursts (dotted lines) parameterized with the light-curve approach described in Sect.~\ref{approach1}. \textbf{a} The 1.1\,mm ($\sim\!280$\,GHz) light curve ($\chi\dmrm{red}^2\!=\!30.5$); \textbf{b} the 37\,GHz light curve ($\chi\dmrm{red}^2\!=\!21.4$); \textbf{c} the 8.0\,GHz light curve ($\chi\dmrm{red}^2\!=\!9.6$).
Each outburst starts simultaneously at all frequencies at the epoch $t_0$ shown in panel \textbf{a}.
The dashed line is the contribution of the jet's hot spot 3C~273A
\label{fit1}
}
\end{figure}

\begin{figure}[tb]
\includegraphics[width=\hsize]{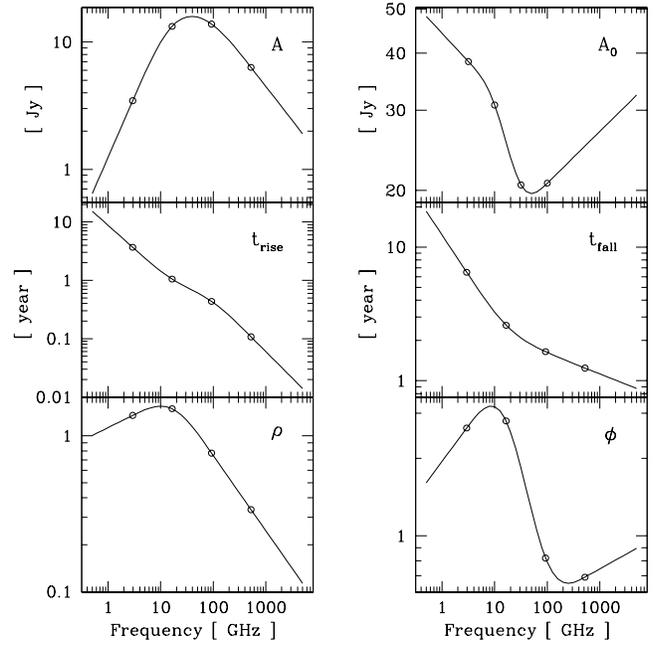}
\caption{
Evolution with frequency of the six parameters: $A(\nu)$, $A_0(\nu)$, $t\dmrm{rise}(\nu)$, $t\dmrm{fall}(\nu)$, $\rho(\nu)$ and $\phi(\nu)$, defined in Sect.~\ref{model1}.
All these functions are a cubic spline passing through the four points with frequencies fixed at $\log{(\nu/\mbox{GHz})}=$ 0.5, 1.0, 1.5 and 2.0 for $A_0(\nu)$ and at $\log{(\nu/\mbox{GHz})}\simeq$ 0.5, 1.25, 2.0 and 2.75 for the five other parameters
\label{param}
}
\end{figure}

\subsection{Results}
\label{results1}

Fig.~\ref{fit1} shows three representative light curves among the twelve fitted simultaneously with the outbursts parameterized as described in Sect.~\ref{model1}.
The major features of the light curves are reproduced by the model with only about one outburst every 1.5 year starting simultaneously at all frequencies.
The overall fit has a reduced $\chi^2$ value of $\chi\dmrm{red}^2\!\equiv\!\chi^2/\mbox{d.o.f.}\!=\!16.1$.
The main discrepancy between the model and the observations arises during 1984--1985, when the very different light curve features in the millimetre and radio domains cannot be correctly described by the 1983.4 flare alone.

The obtained evolution of the parameters with frequency for the typical outburst is shown in Fig.~\ref{param}.
The amplitude $A(\nu)$ of the light curve has a maximum at $\sim\!45$\,GHz.
Both the rise time $t\dmrm{rise}(\nu)$ and the $\mrm{e}$-folding decay time $t\dmrm{fall}(\nu)$ increase monotonically with wavelength.
If we extrapolate the cubic spline to low frequency, it is striking to see that both $t\dmrm{rise}(\nu)$ and $t\dmrm{fall}(\nu)$ tend to very high values of the order of 10 years at 1\,GHz, while the amplitude of the outburst would still be significant ($A(\mbox{1\,GHz})\!\approx\!1$\,Jy).
Due to the lack of submillimetre observations before 1981, the amplitude $A_0(\nu)$ is not constrained at frequencies above $\sim\!300$\,GHz.
The increase of $A_0(\nu)$ at these frequencies -- due to the spline -- is probably not real, but does not affect the fit because the corresponding decay time is short ($t\dmrm{fall}(\mbox{1000\,GHz})\!\approx\!1$ year).
The two exponents $\rho(\nu)$ and $\phi(\nu)$ which describe the shape of the outburst's light curve are both higher at radio frequencies than in the mm/submm domain.
As a consequence, the light curves at higher frequencies have a steeper rise just after the start of the outburst and a steeper decay just after the peak (see Figs.~\ref{func} and \ref{complc}a).

\begin{figure*}[tb]
\includegraphics[width=12cm]{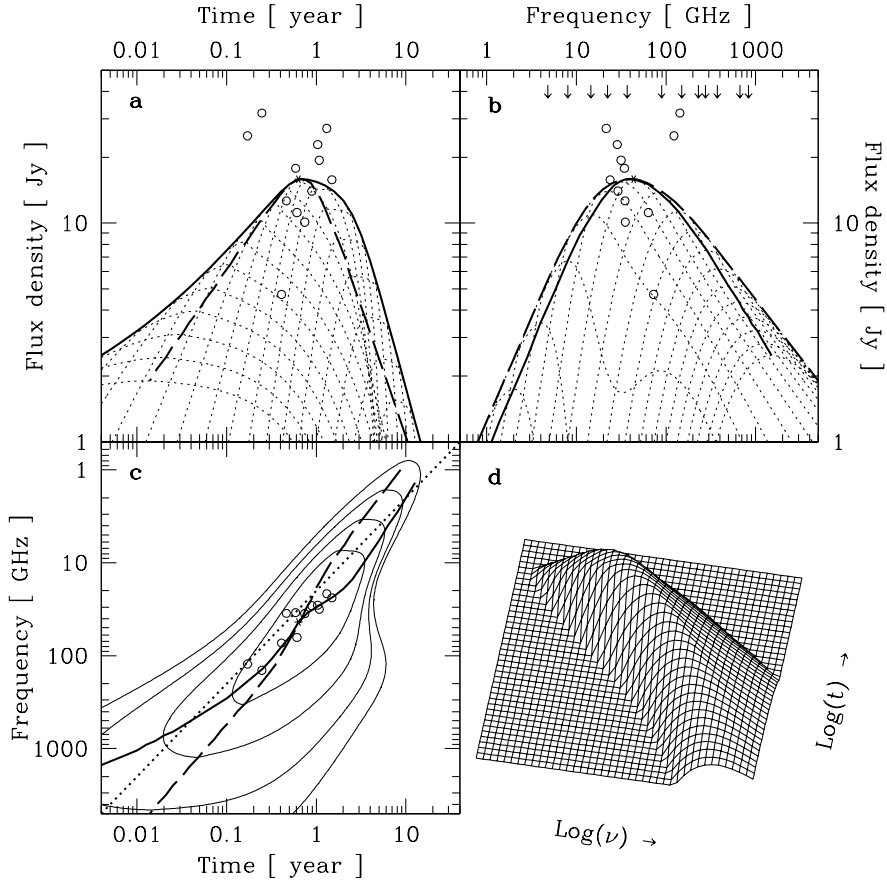}
\hfill
\parbox[b]{55mm}{
\caption{
\textbf{a--d.} Logarithmic spectral and temporal evolution of the typical outburst obtained by the light-curve approach described in Sect.~\ref{approach1}.
Panel \textbf{d} shows the three-dimensional representation in the $(\log{S},\log{\nu},\log{t})$-space.
The other panels show the three Cartesian projections: \textbf{a} light curves at different frequencies spaced by 0.2\,dex; \textbf{b} spectra at different times spaced by 0.2\,dex; \textbf{c} contour plot in the frequency versus time plan.
The thick solid and dashed lines show the evolution of the maximum of the spectra and of the light curves, respectively.
They follow quite well the dotted diagonal in panel \textbf{c} corresponding to $\nu\!\propto\!t^{-1}$.
The star dot shows the point of maximum development of the typical outburst.
The open circles show for each outburst the position that would have this maximum according to the shifts $\Delta\log{S}$, $\Delta\log{\nu}$ and $\Delta\log{t}$.
The arrows in panel \textbf{b} show the frequency distribution of the twelve light curves
\label{evo1}
}
}
\end{figure*}

The five parameters $A(\nu)$, $t\dmrm{rise}(\nu)$, $t\dmrm{fall}(\nu)$, $\rho(\nu)$ and $\phi(\nu)$ define the typical outburst that can be represented in three dimensions in the $(\log{S},\log{\nu},\log{t})$-space as shown in Fig.~\ref{evo1}d.
The three other panels of Fig.~\ref{evo1} show the three Cartesian projections of this surface.
The frequency and time axes cover the same logarithmical range of 4\,dex, so that the dotted diagonal in Fig.~\ref{evo1}c corresponds to $\nu\!\propto\!t^{-1}$.
At least at low frequencies, both the maximum of the spectra and of the light curves follow quite well this diagonal.
The outburst's evolution is thus amazingly symmetric in Figs.~\ref{evo1}a and~\ref{evo1}b.
The maximum amplitude of the typical outburst is of $\sim\!16$\,Jy and is reached after $\sim\!7.5$ months at a frequency of $\sim\!45$\,GHz.
The frequency $\nu\dmrm{m}$ of the spectrum's maximum is steadily decreasing with time (Fig.~\ref{evo1}c).
The corresponding flux density $S\dmrm{m}$ is first increasing with decreasing frequency $\nu\dmrm{m}$ according to $S\dmrm{m}\!\propto\!\nu\dmrm{m}^{-0.7}$, whereas it decreases as $S\dmrm{m}\!\propto\!\nu\dmrm{m}^{+1.0}$ during the final decline of the outburst (Fig.~\ref{evo1}b).
This behaviour corresponds qualitatively to what is expected by shock models (e.g. MG85).

At frequencies above the spectral turnover ($\nu\gg \nu\dmrm{m}$), the spectral index $\alpha$ is first of $\sim\!-0.5$ and steepens very slightly to $\sim\!-0.7$ at the maximum development of the outburst.
The somewhat chaotic behaviour during the final declining phase -- due to the abrupt change in the parameter $\phi(\nu)$ -- does not enable us to define a reasonable spectral index during this last stage.
At frequencies below the spectral turnover ($\nu\ll \nu\dmrm{m}$), the spectral index $\alpha$ is smoothly steepening with time from $\sim\!1.8$ to $\sim\!2.5$.
This is what is expected from a synchrotron source that starts inhomogeneous and progressively becomes homogeneous (e.g. Marscher \cite{M77}).

The twelve individual outbursts have different amplitudes ranging from 5\,Jy up to 32\,Jy for the 1983.0 flare studied by MG85.
The corresponding dispersion $\sigma$ of the amplitude shifts $\Delta\log{S}$ is $\sigma=0.23$, which is slightly smaller than the dispersion of the frequency shifts $\Delta\log{\nu}$ ($\sigma=0.27$) and the time shifts $\Delta\log{t}$ ($\sigma=0.29$).
The amplitude shifts $\Delta\log{S}$ are obviously not correlated with either $\Delta\log{\nu}$ or $\Delta\log{t}$ (Fig.~\ref{evo1}a and b).
This is confirmed by a Spearman rank-order test (Bevington \cite{B69}), which yields that the observed correlations could occur by chance with a probability of more than 60\,\%.
On the contrary, the shifts $\Delta\log{\nu}$ and $\Delta\log{t}$ align well along the $\nu\!\propto\!t^{-1}$ line (Fig.~\ref{evo1}c) and the Spearman's test probability of $<$\,0.01\,\% confirms that this anti-correlation is very significant.

\section{The three-stage approach}
\label{approach2}

In the light-curve approach described above, we model analytically the light curve of an outburst at different frequencies and show that the resulting typical flare is qualitatively in agreement with what is expected by shock models in relativistic jets.
It is thus of interest to derive from the data the parameters that are relevant to those models.

The shock model of MG85 and its generalization by Valtaoja et al. (\cite{VTU92}) describe the evolution of the shock by three distinct stages: 1) a rising phase, 2) a peaking phase and 3) a declining phase\footnote{We use here the terminology introduced by Qian et al. (\cite{QWB96}), because it is purely descriptive and free of any interpretation regarding the physical origin of these stages.}.
The three-stage approach presented below is similar to that of Valtaoja et al. (\cite{VTU92}), in the sense that its aim is simply to qualitatively describe the observations.
It contains however more parameters in order to include those which are relevant to test the physical model of MG85.

The remarks of Sect.~\ref{number} concerning the number of outbursts and the quoted values of the reduced $\chi^2$ apply equally here.

\subsection{Parameterization}
\label{model2}

The self-absorbed synchrotron spectrum emitted by electrons with a power law energy distribution of the form $N(E)\!\propto\!E^{-s}$ can be expressed -- by generalizing the homogeneous case (e.g. Pacholczyk \cite{P70}; Stevens et al. \cite{SLR95}) -- as
\begin{equation}
S_{\nu}=S_1\left(\frac{\nu}{\nu_1}\right)^{\alpha\dmrm{thick}}\frac{1-\exp{(-(\nu/\nu_1)^{\alpha\dmrm{thin}-\alpha\dmrm{thick}})}}{1-\mrm{e}^{-1}}\,,
\label{Snu1}
\end{equation}
where $(\nu/\nu_1)^{\alpha\dmrm{thin}-\alpha\dmrm{thick}}$ is equal to the optical depth $\tau_{\nu}$ at frequency $\nu$.
$S_1$ and $\nu_1$ are respectively the flux density and the frequency corresponding to an optical depth of $\tau_{\nu}\!=\!1$.
At high frequency ($\nu\!\gg\!\nu_1$) the medium is optically thin ($\tau_{\nu}\ll 1$) and the spectrum follows a power law of index $\alpha\dmrm{thin}\!=\!-(s-1)/2$, whereas at low frequency ($\nu\!\ll\!\nu_1$) it is optically thick ($\tau_{\nu}\!\gg\!1$) and the spectral index is $\alpha\dmrm{thick}$.
In the case of a homogeneous source, $\alpha\dmrm{thick}\!=\!+5/2$.

The maximum $S\dmrm{m}\!\equiv\!S_{\nu}(\nu\dmrm{m})$ of the spectrum $S_{\nu}$ is reached at the turnover frequency $\nu\dmrm{m}$ corresponding to an optical depth of $\tau\dmrm{m}\!=\!(\nu\dmrm{m}/\nu_1)^{\alpha\dmrm{thin}-\alpha\dmrm{thick}}$.
$\tau\dmrm{m}$ is obtained by differentiating Eq.~(\ref{Snu1}):
\begin{equation}
\frac{\mrm{d}S_{\nu}}{\mrm{d}\nu}=0\;\Rightarrow\;
\exp{(\tau\dmrm{m})}-1=\left(1-\frac{\alpha\dmrm{thin}}{\alpha\dmrm{thick}}\right)\,\tau\dmrm{m}\,.
\label{taum}
\end{equation}
By developing the exponential of Eq.~(\ref{taum}) to the third order, we obtain a good approximate:
$\tau\dmrm{m}\!=\!\frac{3}{2}\left(\sqrt{1-\frac{8\,\alpha\dmrm{thin}}{3\,\alpha\dmrm{thick}}}-1\right)$.
We can now rewrite Eq.~(\ref{Snu1}) according to the turnover values $\nu\dmrm{m}$, $\tau\dmrm{m}$ and $S\dmrm{m}$ by
\begin{equation}
S_{\nu}=S\dmrm{m}\left(\frac{\nu}{\nu\dmrm{m}}\right)^{\alpha\dmrm{thick}}\frac{1-\exp{(-\tau\dmrm{m}\,(\nu/\nu\dmrm{m})^{\alpha\dmrm{thin}-\alpha\dmrm{thick}})}}{1-\mrm{e}^{-\tau\dmrm{m}}}\,.
\label{Snu2}
\end{equation}

The evolution with time of the self-absorbed synchrotron spectrum of Eq.~(\ref{Snu2}) is assumed to follow three distinct stages: 1) the rising phase for $t\!-\!t_0<t\dmrm{r}$\,; 2) the peaking phase for $t\dmrm{r}\leq t\!-\!t_0\leq t\dmrm{p}$ and 3) the declining phase for $t\!-\!t_0>t\dmrm{p}$.
The subscripts ``$\mrm{r}$'' and ``$\mrm{p}$'' refer to the end of the rising phase and the end of the peaking phase, respectively.
We assume that during each stage $i$ ($i\!=\!1,\,2,\,3$) both the turnover frequency $\nu\dmrm{m}(t)$ and the turnover flux $S\dmrm{m}(t)$ evolve with time as a power law, but with exponents that differ during the three stages:
\begin{equation}
\nu\dmrm{m}(t)\propto t^{\beta_i}\quad\mbox{and}\quad S\dmrm{m}(t)\propto t^{\gamma_i}\quad\Rightarrow\quad S\dmrm{m}\propto \nu\dmrm{m}^{\gamma_i/\beta_i}\,.
\label{temp_evo}
\end{equation}
We thus need ten parameters: $t\dmrm{r}$, $t\dmrm{p}$, $\nu\dmrm{m}(t\dmrm{r})$, $S\dmrm{m}(t\dmrm{r})$, $\beta_1$, $\beta_2$, $\beta_3$, $\gamma_1$, $\gamma_2$ and $\gamma_3$, to describe the evolution of the spectral turnover in the three dimensional $(S,\nu,t)$-space.

The model of MG85 predicts that both the optically thin $\alpha\dmrm{thin}$ and thick $\alpha\dmrm{thick}$ spectral indices should be flatter during the declining phase than during the rising and peaking phases (see Fig.~3 of Marscher et al. \cite{MGT92}).
To test whether the spectrum is actually changing from the rising phase to the declining phase, we allow the two spectral indices $\alpha\dmrm{thin}$ and $\alpha\dmrm{thick}$ to have different values during these two stages.
The transition during the intermediate peaking phase from the values in the rising phase ($\alpha\dmrm{thin}(t\dmrm{r})$ and $\alpha\dmrm{thick}(t\dmrm{r})$) to the values in the declining phase ($\alpha\dmrm{thin}(t\dmrm{p})$ and $\alpha\dmrm{thick}(t\dmrm{p})$) is assumed to be linear with the logarithm of time $\log{(t)}$.
This adds the four parameters $\alpha\dmrm{thin}(t\dmrm{r})$, $\alpha\dmrm{thin}(t\dmrm{p})$, $\alpha\dmrm{thick}(t\dmrm{r})$ and $\alpha\dmrm{thick}(t\dmrm{p})$ to the model, having thus a total of fourteen parameters to fully define the evolution of a typical flare in the $(S,\nu,t)$-space instead of the twenty parameters used in the first approach (Sect.~\ref{model1}).

The specificity of each outburst is modelled with a total of $12\!\times\!4$ parameters exactly as described in Sect.~\ref{model1} for the light-curve approach.
We do not model again the superimposed decays of the outbursts that started before 1979, but simply use the same exponential decay as obtained by the first approach (Sect.~\ref{model1}).
The constant contribution of the jet's hot spot 3C~273A is also considered here.
The total number of parameters in this second parameterization is a bit less than for the first one: 62 ($12\!\times\!4 + 14$) instead of 72.

\begin{figure}[tb]
\includegraphics[width=\hsize]{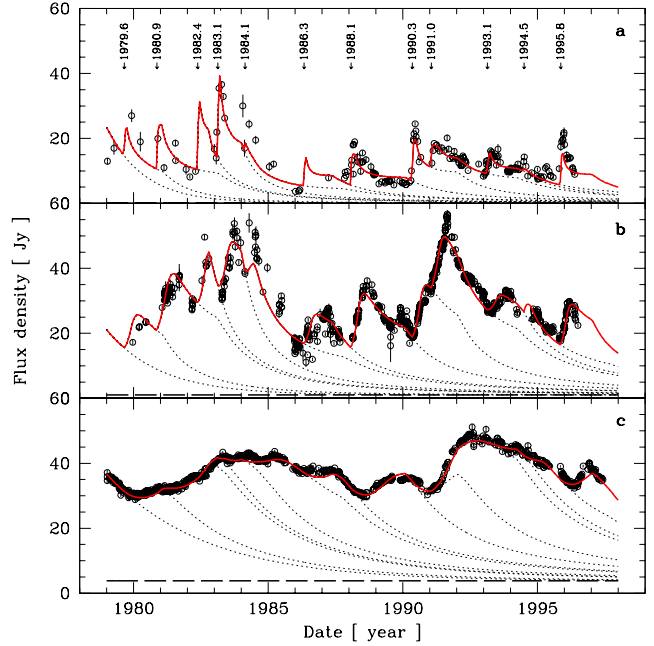}
\caption{
\textbf{a--c.} Same as Fig.~\ref{fit1}, but with the outbursts parameterized according to the three-stage approach described in Sect.~\ref{approach2}. \textbf{a} The 1.1\,mm ($\sim\!280$\,GHz) light curve ($\chi\dmrm{red}^2\!=\!26.8$); \textbf{b} the 37\,GHz light curve ($\chi\dmrm{red}^2\!=\!20.4$); \textbf{c} the 8.0\,GHz light curve ($\chi\dmrm{red}^2\!=\!10.9$)
\label{fit2}
}
\end{figure}

\begin{figure*}[tb]
\includegraphics[width=12cm]{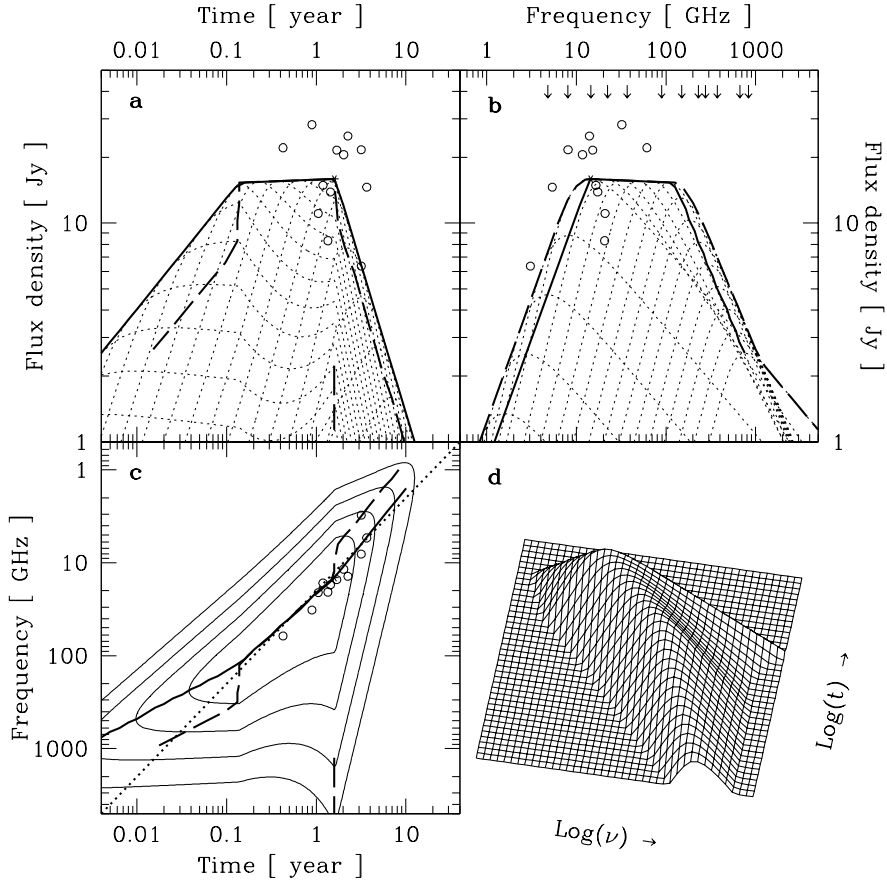}
\hfill
\parbox[b]{55mm}{
\caption{
\textbf{a--d.} Logarithmic spectral and temporal evolution of a typical flare in 3C~273 obtained by the three-stage approach described in Sect.~\ref{approach2}.
The range covered by the axis is the same as in Fig.~\ref{evo1}.
All lines and points are defined as described in Fig.~\ref{evo1}.
The maximum of the spectrum follows the thick solid line.
The three different slopes of this line correspond from left to right in panel \textbf{a} to $\gamma_1$, $\gamma_2$ and $\gamma_3$; in panel \textbf{b} to $\gamma_3/\beta_3$, $\gamma_2/\beta_2$ and $\gamma_1/\beta_1$; and in panel \textbf{c} to $\beta_1$, $\beta_2$ and $\beta_3$.
The values of these slopes are given in Table~\ref{tab1}.
Notice in panel \textbf{b} the change of the optically thin spectral index $\alpha\dmrm{thin}$ from $-1.1$ during the rising phase to $-0.5$ during the declining phase.
This change is responsible for the strange shapes of the high frequency light curves in panel \textbf{a}
\label{evo2}
}
}
\end{figure*}

\subsection{Results}
\label{results2}

To allow a better comparison with the results of the first approach (Sect.~\ref{model1}), we show in Fig.~\ref{fit2} the same light curves as in Fig.~\ref{fit1}.
The reduced $\chi^2$ of the overall fit is now of $\chi\dmrm{red}^2\!=\!17.8$.
The higher frequency light curves are relatively better described here than with the first approach (compare Figs.~\ref{fit1} and \ref{fit2}).
The start times $t_0$ of the outbursts are very similar to those obtained by the first approach, except for the fourth flare which is now starting much later at $t_0$\,=\,1984.1 instead of 1983.4.
This later $t_0$ seems to be in better agreement with the observations, but the behaviour of 3C~273 during 1984--1985 is still poorly described.

\begin{table}[tb]
\caption{
Values of the parameters defined in Sect.~\ref{model2} corresponding to the evolution of the typical outburst shown in Fig.~\ref{evo2}.
The two first columns display the fourteen best fit parameters, whereas other related parameters are shown in the last column
}
\label{tab1}
\begin{flushleft}
\begin{tabular}{@{}l@{ }r|l@{ }r|lr@{}}
\hline
\rule[-0.7em]{0pt}{2.0em}Param.& Value& Param.& Value& Param.& Value\\
\hline
\rule{0pt}{1.2em}$t\dmrm{r}$& 0.14\,year& $t\dmrm{p}$& 1.63\,year& & \\
$\nu\dmrm{m}(t\dmrm{r})$& 120\,GHz& & & $\nu\dmrm{m}(t\dmrm{p})$& 13.8\,GHz\\
$S\dmrm{m}(t\dmrm{r})$& 15.3\,Jy& & & $S\dmrm{m}(t\dmrm{p})$& 15.9\,Jy\\
$\alpha\dmrm{thin}(t\dmrm{r})$& $-$1.09& $\alpha\dmrm{thin}(t\dmrm{p})$& $-$0.48& & \\
$\alpha\dmrm{thick}(t\dmrm{r})$& $+$1.55& $\alpha\dmrm{thick}(t\dmrm{p})$& $+$1.74& & \\
$\beta_1$& $-$0.51& $\gamma_1$& $+$0.51& $\gamma_1/\beta_1$& $-$0.99\\
$\beta_2$& $-$0.88& $\gamma_2$& $+$0.02& $\gamma_2/\beta_2$& $-$0.02\\
$\beta_3$& $-$1.19& $\gamma_3$& $-$1.36& $\gamma_3/\beta_3$& $+$1.14\\
\hline
\end{tabular}
\end{flushleft}
\end{table}

The obtained values of the parameters are given in Table~\ref{tab1}.
They correspond to the spectral and temporal evolution of the typical outburst shown in Fig.~\ref{evo2}.
If the tracks followed by the maximum of the spectra and of the light curves are similar to those obtained by the first approach (Fig.~\ref{evo1}), the spectral evolution of the outburst derived here is quite different.
We obtain that the spectral turnover flux $S\dmrm{m}$ increases during the first 50 days ($t\dmrm{r}\!=\!0.14$ year) with decreasing turnover frequency $\nu\dmrm{m}$ as $S\dmrm{m}\!\propto\!\nu\dmrm{m}^{-1.0}$.
The subsequent very flat peaking phase is found to be relatively long, since it lasts 1.5 year and spans nearly one order of magnitude in frequency from 120\,GHz to 13.8\,GHz.
The final declining phase is quite abrupt with a relation between $S\dmrm{m}$ and $\nu\dmrm{m}$ of $S\dmrm{m}\!\propto\!\nu\dmrm{m}^{+1.1}$.
The optically thin spectral index $\alpha\dmrm{thin}$ is found to be clearly steeper in the rising phase than in the declining phase.
It is flattening by $\Delta\alpha\dmrm{thin}\!=\!+0.6$ during the peaking phase from $\alpha\dmrm{thin}(t\dmrm{r})\!=\!-1.1$ to $\alpha\dmrm{thin}(t\dmrm{p})\!=\!-0.5$.
The optically thick spectral index $\alpha\dmrm{thick}$ is found to be more constant with a slight tendency to steepen with time.
It has a mean value of $\alpha\dmrm{thick}\!=\!+1.65$ and is steepening by $\Delta\alpha\dmrm{thick}\!=\!+0.2$ during the peaking phase.

For each outburst we obtain logarithmic shifts in amplitude $\Delta\log{S}$, frequency $\Delta\log{\nu}$ and time $\Delta\log{t}$, which are similar to those obtained by the first approach (Sect.~\ref{approach1}).
The dispersions $\sigma$ of $\Delta\log{S}$, $\Delta\log{\nu}$ and $\Delta\log{t}$ are $0.20$, $0.34$ and $0.27$, respectively.
A possible correlation of $\Delta\log{S}$ with either $\Delta\log{\nu}$ or $\Delta\log{t}$ is again not significant: the Spearman's test probability that stronger correlations could occur by chance is $>$\,40\,\%.
On the contrary, the strong correlation observed between $\Delta\log{\nu}$ and $\Delta\log{t}$ is most probably real (Spearman's test probability $<10^{-6}$).

\section{Discussion}
\label{discussion}

\begin{figure}[tb]
\includegraphics[width=\hsize]{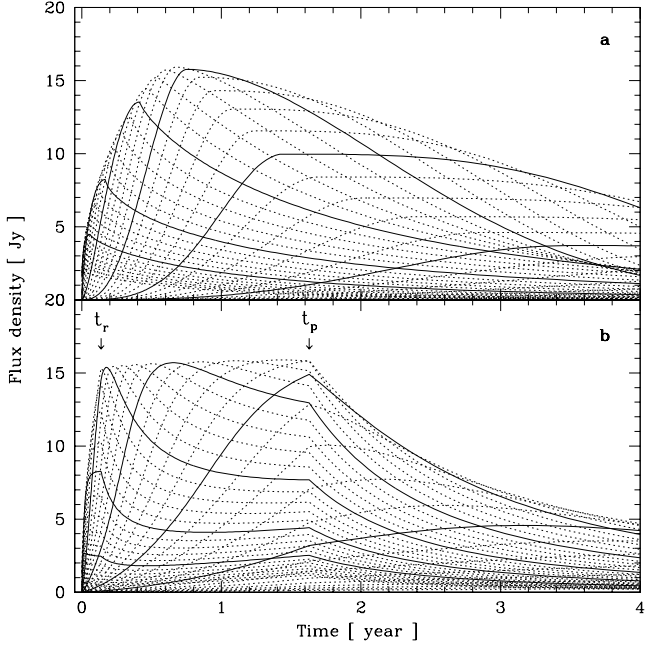}
\caption{
\textbf{a and b.} Light curves at different frequencies for the typical outburst obtained with the light-curve approach (\textbf{a}) and the three-stage approach (\textbf{b}).
The dotted light curves are spaced by 0.1\,dex in frequency and the six solid light curves are at frequencies $\log{(\nu/\mbox{GHz})}\!=\!3.0,\,2.5,\ldots,\,0.5$, in order of increasing time scales
\label{complc}
}
\end{figure}

The two approaches presented above give comparable results, but differ concerning the existence of a nearly constant peaking phase and the shapes of the spectra (compare Figs.~\ref{evo1}b and \ref{evo2}b).
The origin of these differences can be understood by comparing the light curve profiles obtained by the two approaches, which are shown in Fig.~\ref{complc}.
It is clear that the first approach allowing only a rising phase and a declining phase cannot mimic the three-stage profiles of Fig.~\ref{complc}b resulting from the second approach (Sect.~\ref{model2}).
On the other hand, only the light-curve approach is able to produce a round peaking phase as seen in Fig.~\ref{complc}a.
In a forthcoming paper (T\"urler et al. in preparation), we will present the results of an hybrid approach, which incorporates the advantages of both approaches in order to better define the properties of the typical outburst.

\subsection{Do the outbursts correspond to VLBI components?}
\label{vlbi}

The decomposition of the light curves into distinct outbursts was motivated by the observation with very long baseline interferometry (VLBI) of distinct components in the jet structure of 3C~273.
Since the detection of a new VLBI component (Krichbaum et al. \cite{KBK90}) associated with the strong optical/infrared flare of 1988 in 3C~273 (Courvoisier et al. \cite{CRB88}), there is good evidence that outbursts are related to the ejection of new VLBI knots.
To test whether all outbursts are actually associated with superluminal components, we compare in Table~\ref{tab2} the start time $t_0$ of an outburst -- as obtained by the three-stage approach -- with the ejection time ``$t_0$ (knot)'' of a new VLBI knot as given by  Abraham et al. (\cite{ACZ96}) and Zensus et al. (\cite{ZUC90}).
For each of the eight first outbursts, we can identify one or two possibly associated VLBI components.

To test further this relationship, we compare the flux densities ``$F\dmrm{obs}$ (knot)'' of the VLBI components observed at epoch $t\!=\!1991.15$ and at a frequency of 10.7\,GHz (Abraham et al. \cite{ACZ96}) with the flux densities $F\dmrm{exp}(t\!=\!1991.15,\,\nu\!=\!10.7\,\mrm{GHz})$ expected at the same epoch and the same frequency according to the outburst parameters derived here.
Table~\ref{tab2} shows that for the five first outbursts there is always one of the possibly associated knots (indicated by an arrow), which has the expected flux.
For the three remaining outbursts and especially for the 1988.1 flare, the relation between $F\dmrm{obs}$ (knot) and $F\dmrm{exp}$ is not obvious.
At this epoch however, the possibly associated components are still strongly blended by the core emission (component ``D'') or might even still be part of the unresolved core\footnote{In our model the core emission is entirely due to a superimposition of outbursts unresolved by the VLBI.}.
The total flux $F\dmrm{exp}\!=\!20.9$\,Jy expected by the 1986.3, 1988.1 and 1990.3 outbursts is indeed equal to the observed total flux $F\dmrm{obs}\!=\!21.4\pm 0.7$\,Jy of the C9, C10 and D components.
These results strongly suggest that there is a close relation between the outbursts and the VLBI knots and hence that our decomposition describes a real physical aspect of the jet.

\begin{table}[tb]
\caption{
Relation between VLBI components and the eight first outbursts as obtained with the three-stage approach (Sect.~\ref{approach2}).
The parameters in this table are defined in Sect.~\ref{vlbi}
}
\label{tab2}
\begin{flushleft}
\begin{tabular}{@{}llcrr@{~~}r@{}}
\hline
\rule[-0.7em]{0pt}{2.0em}$t_0$& Knot& $t_0$ (knot)& \multicolumn{1}{c}{$F\dmrm{exp}$}& \multicolumn{1}{c}{$F\dmrm{obs}$ (knot)}\\
\hline
\rule{0pt}{1.2em}1979.6& C5& 1978.6$\,\pm\,$0.04& 0.4\,Jy& 0.7$\,\pm\,$0.4\,Jy& $\leftarrow$\\
 & C6& 1980.0$\,\pm\,$0.04& & 1.9$\,\pm\,$0.5\,Jy& \\
1980.9& C6& 1980.0$\,\pm\,$0.04& 1.9\,Jy& 1.9$\,\pm\,$0.5\,Jy& $\leftarrow$\\
 & X& $<$\,1985.6& & $<$\,0.5\,Jy& \\
1982.4& C7& 1982.2$\,\pm\,$0.4& 0.2\,Jy& $<$\,0.5\,Jy& $\leftarrow$\\
1983.1& C7a& 1983.1$\,\pm\,$0.00& 0.9\,Jy& $<$\,0.5\,Jy& \\
 & C7b& 1983.6$\,\pm\,$0.09& & 1.2$\,\pm\,$0.3\,Jy& $\leftarrow$\\
1984.1& C7b& 1983.6$\,\pm\,$0.09& 3.8\,Jy& 1.2$\,\pm\,$0.3\,Jy& \\
 & C8& 1984.7$\,\pm\,$0.10& & 4.4$\,\pm\,$0.5\,Jy& $\leftarrow$\\
1986.3& Cx& $<$\,1988.2& 1.0\,Jy& $<$\,0.5\,Jy& \\
1988.1& C9& 1988.4$\,\pm\,$0.17& 11.2\,Jy& 1.8$\,\pm\,$0.4\,Jy& \\
1990.3& C10& $<$\,1990.2& 8.7\,Jy& 6.4$\,\pm\,$0.5\,Jy& \\
 & D& & & 13.2$\,\pm\,$0.3\,Jy& \\
\hline
\end{tabular}
\end{flushleft}
\end{table}

\subsection{How can we understand the peculiarities of individual outbursts?}

The relation found between the outbursts and the VLBI knots (Sect.~\ref{vlbi}) has established that our decomposition is not purely mathematical, but does correspond to a physical reality.
There should therefore be a physical origin to the clear anti-correlation found between the frequency shifts $\Delta\log{\nu}$ and the time shifts $\Delta\log{t}$ of the individual outbursts.
The observed frequency shifts $\Delta\log{\nu}$ confirm that 3C~273 emits both low- and high-frequency peaking outbursts (Lainela et al. \cite{LVT92}).
The relation between $\Delta\log{\nu}$ and $\Delta\log{t}$ clearly shows that high-frequency peaking flares evolve faster than low-frequency peaking outbursts.
The alignment of the shifts along the $\nu\!\propto\! t^{-1}$ line (Figs.~\ref{evo1}c and \ref{evo2}c) further suggests the relation $\Delta\log{\nu}\!=\!-\Delta\log{t}$.

The origin of this relation could be due to a change $\Delta\log{\mathcal{D}}$ of the Doppler factor $\mathcal{D}=\gamma^{-1}(1-\beta\cos{\theta})^{-1}$, which depends on the flow speed $\beta\!=\!v/c$, the Lorentz factor $\gamma\!=\!(1-\beta^{2})^{-1/2}$ and the angle to the line of sight $\theta$.
Observed quantities (unprimed) are related to emitted quantities (primed) as (e.g. Hughes \& Miller \cite{HM91}; Pearson \& Zensus \cite{PZ87}):
\begin{eqnarray}
\nu=\mathcal{D}\,\nu^{\prime}& \Rightarrow& \Delta\log{\nu}=\Delta\log{\mathcal{D}}+\Delta\log{\nu^{\prime}}\\
t=\mathcal{D}^{-1}\,t^{\prime}& \Rightarrow& \Delta\log{t}=-\Delta\log{\mathcal{D}}+\Delta\log{t^{\prime}}\\
S(\nu)=\mathcal{D}^{3}\,S^{\prime}(\nu^{\prime})& \Rightarrow& \Delta\log{S}=3\,\Delta\log{\mathcal{D}}+\Delta\log{S^{\prime}}
\end{eqnarray}
If we assume that in the jet frame all outbursts are alike (i.e. $\Delta\log{k^{\prime}}\!=\!0,\,\forall\,k\!=\!S,\,\nu,\,t$), the observed relation $\Delta\log{\nu}\!=\!-\Delta\log{t}$ can be interpreted as a change $\Delta\log{\mathcal{D}}$ of the Doppler factor from one outburst to the other.
In this case, however, there should also be correlations between $\Delta\log{S}$ and both $\Delta\log{\nu}$ and $\Delta\log{t}$, which are not observed.

\begin{figure}[tb]
\includegraphics[width=\hsize]{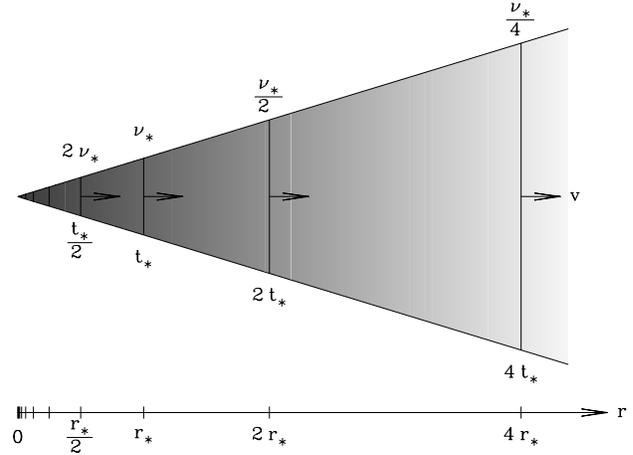}
\caption{
Schematic representation of a conical jet with constant speed $v$ in which the frequency of maximum emission $\nu\dmrm{m}$ (greyscale) is inversely proportional to the distance down the jet $r\!=\!v\,t$ ($\nu\dmrm{m}\!\propto\!r^{-1}\propto\!t^{-1}$).
Let us assume that the peaking phase of an outburst corresponds to an octave of radius ($r\rightarrow 2\,r$).
This phase has a duration and a frequency range which clearly depend on the distance $r_0$ at which the shock forms.
Inner outbursts peaking from $r_{\ast}/2$ to $r_{\ast}$ are four times shorter than outer outbursts peaking from $2\,r_{\ast}$ to $4\,r_{\ast}$, while their emission is maximum at four times higher frequencies
\label{jetfig}
}
\end{figure}

Alternatively, we can consider that the Doppler factor does not change ($\Delta\log{\mathcal{D}}\!=\!0$) and that the observed relation between $\Delta\log{\nu}$ and $\Delta\log{t}$ is intrinsic and independent of possible flux variations $\Delta\log{S}$.
Such a correlation might be related to the distance from the core at which the shock forms (Lainela et al. \cite{LVT92}).
Indeed, Blandford (\cite{B90}) shows that for a simple conical jet with constant speed $v$ the frequency of maximum emission $\nu\dmrm{m}$ is inversely proportional to the distance down the jet $r\!=\!v\,t$ ($\nu\dmrm{m}\!\propto\!r^{-1}$), while the corresponding flux density $S\dmrm{m}$ is constant.
Since the speed $v$ is constant, the turnover frequency $\nu\dmrm{m}$ is then also inversely proportional to time ($\nu\dmrm{m}\!\propto\!t^{-1}$), as observed.
If a shock forms in such an underlying jet at a distance $r_0$ from the core, both the frequency range of the emission and the time scale of the evolution will depend on the distance $r_0$, as illustrated in Fig.~\ref{jetfig}.
We therefore propose that short-lived and high-frequency peaking flares are actually \textit{inner} outbursts, whereas long-lived and low-frequency peaking flares are \textit{outer} outbursts.

This interpretation is supported by the existence of short-lived VLBI components which are only seen close to the core.
In our decomposition, the two most short-lived and the most high-frequency peaking outbursts are the two successive flares of 1982.4 and 1983.1.
Their start times correspond well to the period from 1981 to 1983 during which only short-lived VLBI components were formed (Abraham et al. \cite{ACZ96}).
If our interpretation is right, the shifts $\Delta\log{\nu}$ and $\Delta\log{t}$ that we obtain suggest that the 1982.4 flare would have formed about two times closer to the core than the 1983.1 flare and four times closer than the typical outburst.

\subsection{What are the constraints for shock models?}

According to the shock model of MG85, the optically thin spectral index $\alpha\dmrm{thin}$ should be steeper during the two first stages of the outburst evolution than the usual value of $\alpha\dmrm{thin}\!=\!-(s-1)/2$ (Sect.~\ref{model2}).
A steeper index arises due to the fact that the thickness $x$ of the emitting region behind the shock front is proportional to the cooling time $t\dmrm{cool}$ of the electrons suffering radiative (Compton and/or synchrotron) losses.
During the rising and peaking phases, radiative losses are dominant and therefore the thickness $x$ is frequency dependent as $x\!\propto\!t\dmrm{cool}\!\propto\!\nu^{-1/2}$, which leads to a steeper optically thin spectral index of $\alpha\dmrm{thin}\!=\!-s/2$.
Until now, the expected flattening of the spectral index by $\Delta\alpha\dmrm{thin}\!=\!+0.5$ from the rising and peaking phases to the declining phase was never observed and furthermore the optically thin spectral index $\alpha\dmrm{thin}$ observed at the beginning of the outburst was often found to be already too flat ($\alpha\dmrm{thin}\!>\!-1/2$) to allow the expected subsequent flattening (Valtaoja et al. \cite{VHL88}; Lainela \cite{L94}).

The present result that the optically thin spectral index is flattening with time by $\Delta\alpha\dmrm{thin}\!=\!+0.6$ is in good agreement with the change of $\Delta\alpha\dmrm{thin}\!=\!+0.5$ expected by the shock model of MG85.
The observed flattening of the spectrum is contrary to the steepening with time expected as a result of radiative energy losses by the electrons.
The observed behaviour can however also be understood as a change of slope with frequency rather than with time and thus it could conceivably be due to a spectral break that steepens the optically thin spectral index by a factor of 0.5 at higher frequencies.
Such a break is expected in the case of continuous injection or reacceleration of electrons suffering radiative losses (Kardashev \cite{K62}) and is observed in several hot spots including 3C~273A (Meisenheimer et al. \cite{MRH89}).
Whatever the interpretation, the flatter index, $\alpha\dmrm{thin}(t\dmrm{p})$, is the relevant index to determine that the electron energy index $s$ ($N(E)\!\propto\!E^{-s}$) is $s\!=\!1-2\,\alpha\dmrm{thin}(t\dmrm{p})\!=\!+2.0$.
This value corresponds to the average value observed in several hot spots (Meisenheimer et al. \cite{MRH89}) and is in agreement with the values expected if the electrons are accelerated by a Fermi mechanism in a relativistic shock (e.g. Longair \cite{L94b}).

The long flat peaking phase observed in 3C~273 contrasts with the complete absence of this stage in 3C~345 (Stevens et al. \cite{SLR96}).
This difference is surprising, because the outburst's evolution is otherwise very similar in these two objects with nearly the same indices for the rising and the declining phases: $\gamma_1/\beta_1\!=\!-0.99$ in 3C~273 and $-0.86$ in 3C~345 and $\gamma_3/\beta_3\!=\!+1.14$ in 3C~273 and $+0.98$ in 3C~345 ($S\dmrm{m}\!\propto\!\nu\dmrm{m}^{\gamma_i/\beta_i}$).
A value of $\gamma_3/\beta_3\!\sim\!+1$ was also found in several other sources by Valtaoja et al. (\cite{VHL88}).
This decrease of the turnover flux with decreasing frequency is steeper than expected by the simplest model of MG85; i.e. with a conical adiabatic jet having a constant Doppler factor $\mathcal{D}$.
With $s\!=\!2$ and a magnetic field $B$ oriented perpendicular to the jet axis, their model predicts $\gamma_3/\beta_3\!=\!+0.45$.
This discrepancy between the observations and the shock model of MG85 was already pointed out by Stevens et al. (\cite{SLR96}).
We refer the reader to their discussion of two more general cases of the MG85 model: 1) a straight non-adiabatic jet and 2) a curved adiabatic jet.
With the observed values of the indices $\beta_3$ and $\gamma_3$, these authors could determine the two free parameters of the model.
In our case, with the constraints of all six indices $\beta_i$ and $\gamma_i$ ($i\!=\!1,\,2,\,3$), we could not find a good agreement with either of the two models mentioned above.
In a forthcoming paper (T\"urler et al. in preparation), we will further discuss this point and explore whether a non-conical non-adiabatic curved jet can well describe the observations.

\section{Summary and conclusion}
\label{summary}

By using most available submillimetre-to-radio observations of 3C~273, we have been able to extract the properties of the spectral and temporal evolution of a typical outburst.
The new approach we defined consists in decomposing the light curves into several self-similar outbursts.
The main results of our decomposition are the followings:
\begin{itemize}
\item It is possible to understand the very different shapes of the submillimetre-to-radio light curves of 3C~273 with only about one outburst every 1.5 year starting simultaneously at all frequencies.
\item There is no need to invoke any underlying quiescent emission apart from the weak contribution of the jet's hot spot 3C~273A.
\item The outbursts that we identify do well correspond to the observed VLBI components in the jet.
\item There is good evidence that short-lived and high-frequency peaking flares are emitted closer to the core of the jet than long-lived and low-frequency peaking outbursts.
\item The spectral and temporal evolution of the outbursts is found to be in good qualitative agreement with the evolution expected by shock models in relativistic jets.
\item We observe a flattening of the optically thin spectral index from the rising to the declining phase of the shock evolution, which supports the idea proposed by MG85 that radiative (synchrotron and/or Compton) losses are the main cooling process of the electrons during the initial phase of the outburst.
\end{itemize}

We are aware that our decomposition is far from describing the detailed structure of the light curves and that the jet emission is much more complicated than this work tries to show.
Nevertheless, the results suggest that the outbursts we identified are closely related to the VLBI knots, and hence that they describe a physical aspect of the jet.
The new approach presented here is a powerful tool to derive the observed properties of millimetre and radio outbursts.
It allows comparison between shock models and the observations and we are confident that such decompositions are able to further constrain present and future shock models.
Finally, we would like to stress the importance of long-term multi-wavelength monitoring campaigns, which turn out to be essential towards a better understanding of the physics involved in relativistic jets.

\end{document}